\documentclass{article}
\usepackage{spconf,amsmath,graphicx,hyperref}

\usepackage{spconf,amsmath,graphicx}
\usepackage{multirow}
\usepackage{booktabs}
\usepackage{url}
\usepackage{subcaption}
\usepackage{siunitx}
\usepackage{pgfplots}
\usepackage{algorithm}
\usepackage{algpseudocode}
\usepackage{tikz}
\usepackage{algorithm}      
\usepackage{algpseudocode}  
\usepackage{amsmath}  
\usetikzlibrary{shapes, arrows, positioning}
\usepackage{fontawesome5} 
\pgfplotsset{compat=1.18}
\title{ArtiFree: Detecting and Reducing Generative Artifacts in Diffusion-based Speech Enhancement}

\name{Bhawana Chhaglani$^{*}$\thanks{$^{*}$ This work was done when the author was on an internship at Meta}, Yang Gao$^{+}$, Julius Richter$^{\pm}$, Xilin Li$^{+}$, Syavosh Zadissa$^{+}$, Tarun Pruthi$^{+}$, Andrew Lovitt$^{+}$}
\address{$^{*}$University of Massachusetts Amherst, $^{+}$Meta Reality Labs, $^{\pm}$Meta Superintelligence Labs}

\begin{document}
\ninept
\maketitle
\begin{abstract}
Diffusion-based speech enhancement (SE) achieves natural-sounding speech and strong generalization, yet suffers from key limitations like generative artifacts and high inference latency. 
In this work, we systematically study artifact prediction and reduction in diffusion-based SE. We show that variance in speech embeddings can be used to predict phonetic errors during inference.
Building on these findings, we propose an ensemble inference method guided by semantic consistency across multiple diffusion runs. This technique reduces WER by 15\% in low-SNR conditions, effectively improving phonetic accuracy and semantic plausibility. Finally, we analyze the effect of the number of diffusion steps, showing that adaptive diffusion steps balance artifact suppression and latency. Our findings highlight semantic priors as a powerful tool to guide generative SE toward artifact-free outputs.
\end{abstract}


\section{Introduction}
\label{sec:intro}
Speech enhancement (SE) improves intelligibility and quality by suppressing noise, enabling robust applications in telephony, transcription, and assistive hearing devices. Generative models, particularly diffusion-based approaches such as score-based generative models for speech enhancement (SGMSE)~\cite{welker2022speech,richter2023speech} and Schr\"odinger Bridge (SB)~\cite{jukic2024schr,richter2025investigating}, have emerged as powerful alternatives to predictive methods (e.g., CRNN \cite{yin2020phasen,hu2020dccrn}, SEMamba \cite{chao2024investigation}). By learning the distribution of clean speech given noisy speech, diffusion-based SE can produce natural-sounding results and generalize to unseen noise conditions.
However, few challenges remain. First, diffusion models often introduce \emph{generative artifacts}, including phoneme insertions/substitutions, hiss, breathing artifacts, and high-frequency distortions, which degrade automatic speech recognition (ASR) performance despite favorable perceptual scores. 
Second, the output of these models is inherently stochastic, as it is sampled from the posterior distribution of clean speech given noisy speech, leading to natural variability across runs.
Lastly, inference requires multiple reverse diffusion steps, leading to high \emph{latency} unsuitable for real-time deployment.

In this work, we ask: \emph{How can we detect and reduce artifacts in diffusion-based SE, while maintaining efficiency?} 
Unlike predictive SE models, which regress to the mean of the posterior under an MSE objective and thus produce a single deterministic output, diffusion SE models allow sampling multiple plausible outputs from the posterior. However, not all sampled outputs are semantically correct: some contain hallucinated or substituted phonemes. This raises the need for strategies to \emph{select} the most semantically consistent sample rather than averaging across them. Inspired by observations in StoRM \cite{lemercier2023storm}, we argue that while generative sampling provides diversity, semantic priors are required to steer outputs toward artifact-free speech. Our work proposes such a strategy through semantic consistency-based ensemble inference.
This strategy confirms that the model generates optimal enhanced output by selecting the output with the most consistent speech embeddings out of multiple sampling runs. 

\noindent\textbf{Contributions.} 
This paper makes the following contributions. 
(1) We explore intrusive and non-intrusive metrics for artifact detection, demonstrating that PESQ/STOI fail to capture phoneme-level errors, and propose variance of speech embeddings across multiple samples as an inference-time indicator of artifact-prone regions. 
(2) We introduce an ensemble inference method guided by semantic consistency, which reduces phoneme artifacts and achieves up to 15\% relative word error rate (WER) improvement in low-SNR conditions. 
(3) We analyze the trade-off between artifact reduction and latency, showing that while ensemble inference increases real-time factor (RTF), reducing the number of diffusion steps ($N$) provides a favorable quality–efficiency balance. Specifically, the adaptive $N$ approach yields WER improvements with a 27\% reduction in RTF, at the cost of a slight 0.1 drop in PESQ. 
Together, these results provide new insights into the nature of generative artifacts in diffusion SE and propose practical strategies to detect and reduce them in high fidelity applications. We will make the Source code and artifacts test set available.
\section{Background}
\label{sec:background}

Diffusion-based generative models have recently advanced SE by learning distributions over clean speech and progressively denoising noisy inputs \cite{welker2022speech,richter2023speech,richter2025investigating,de2023behavior}. SGMSE~\cite{welker2022speech} solves a stochastic differential equation with a learned score network, while the Schr\"odinger Bridge (SB)~\cite{jukic2024schr,richter2025investigating} casts speech enhancement as an optimal transport problem. These approaches often outperform predictive baselines in terms of perceptual quality and robustness \cite{richter2023speech,lemercier2025diffusion}.

\begin{figure}
    \centering
    \includegraphics[width=0.6\linewidth]{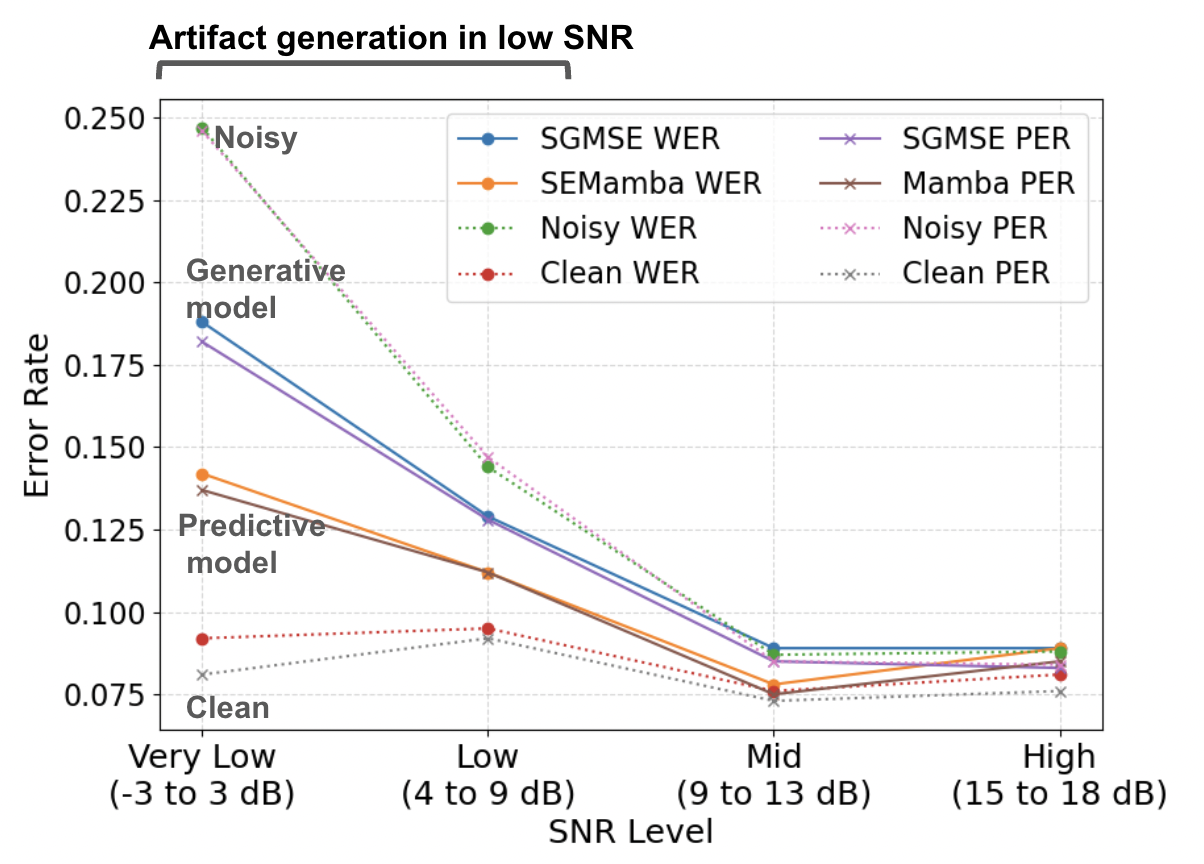}
    \caption{ASR results of generative and predictive models in different SNR levels of VoiceBank-DEMAND test set}
    \label{fig:asr}
\vspace{-0.2cm}
\end{figure}

A key limitation, however, is the emergence of \textit{generative artifacts}. Unlike predictive models, which mainly distort or suppress existing speech, diffusion-based SE can ``hallucinate'' new content. Artifacts include phonetic errors such as insertions or substitutions, spurious breathing or hissing, robotic tones, and high-frequency attenuation \cite{lemercier2025diffusion}. These effects are most pronounced at low SNR as shown in Figure \ref{fig:asr}, where uncertainty drives the model to generate plausible but incorrect phonetic structures, leading to poor ASR performance despite high PESQ or STOI scores \cite{de2023behavior}. Existing metrics fail to fully capture these errors: intrusive metrics emphasize energy-based distortions at the signal-level, while non-intrusive predictors favor naturalness and overrate generative outputs. Complementary measures such as Levenshtein phoneme distance (LPD) and hallucination error rate (HER) have been proposed to address this gap~\cite{pirklbauer2023evaluation,atwany2025lost}.

Efforts to reduce artifacts in SE have largely focused on predictive models, e.g., loss functions penalizing distortions~\cite{guan2024reducing} or no-reference detectors for hum and hiss~\cite{higham2022no}. For generative SE, only limited attempts exist: two-stage pipelines such as StoRM~\cite{lemercier2023storm}, conditioning strategies like SBSE~\cite{wang2024diffusion}, and training- or sampling-time regularization~\cite{tai2023dose,emura2024estimation}. Yet, phoneme hallucinations and spurious vocalizations remain unsolved, motivating systematic frameworks for artifact detection and reduction in diffusion SE.

\section{Method}
\label{sec:methodology}
\begin{figure}[t]
\centering
\begin{tikzpicture}
\begin{axis}[
    ybar,
    width=0.65\linewidth,   
    height=3.2cm,           
    bar width=4pt,          
    ylabel={Correlation},
    symbolic x coords={LSD, Wav2vec dist., VAD, Formant, Wav2vec Confidence, SNR},
    xtick=data,
    x tick label style={rotate=30, anchor=east, font=\scriptsize},
    ymin=-1, ymax=1,
    legend style={at={(0.5,1.05)},anchor=south,legend columns=2, font=\scriptsize},
    tick label style={font=\scriptsize},
    ylabel style={font=\scriptsize}
]
\addplot+[fill=blue!60] coordinates {(LSD,-0.85) (Wav2vec dist.,-0.61) (VAD,-0.43) (Formant,-0.10) (Wav2vec Confidence,-0.13) (SNR,0.62)};
\addplot+[fill=orange!80] coordinates {(LSD,-0.74) (Wav2vec dist.,-0.53) (VAD,-0.51) (Formant,-0.12) (Wav2vec Confidence,-0.14) (SNR,0.56)};
\legend{PESQ, STOI}
\end{axis}
\end{tikzpicture}
\caption{\small Correlation of artifact metrics with PESQ and STOI.}
\label{fig:correlation_bar}
\vspace{-0.2cm}
\end{figure}
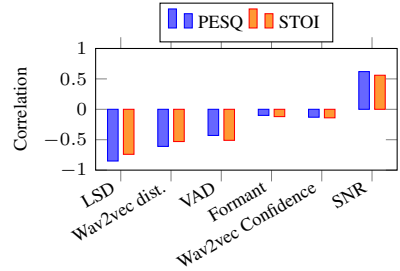

In this section, we describe the methodology of the proposed SE system called \textit{ArtiFree}. We investigate intrusive and non-intrusive metrics for detecting and quantifying generative artifacts.
We compute log spectral distance (LSD), speech embedding distance (wav2vec cosine distance), Voice Activity Detection (VAD) mismatch duration, formants bandwidth divergence between clean and enhanced audio files. We also explore non-intrusive metrics like wav2vec confidence score of the enhanced audio file. Using manual listening test and spectral inspection, we tagged artifact files and non-artifact files and observed that LSD is sensitive to high-frequency attenuation and hiss sounds artifacts, and can also capture phoneme errors. Wav2vec cosine distance uniquely captured phoneme insertions, substitutions or deletions as it directly encodes phoneme-level similarity. VAD mismatch depends highly on tuning the VAD parameters and can sometimes detect breathing or click artifacts. Formant bandwidth divergence failed to detect some phoneme artifacts. So, we found for phoneme artifacts, wav2vec embedding distance is the most suitable metric.\\
\noindent\textbf{Correlation Analysis}: While analyzing correlation between artifact metrics and speech quality metrics on VoiceBank test set (Figure \ref{fig:correlation_bar}), we found that wav2vec cosine distance is only moderately correlated with pesq (-0.61) and stoi (-0.53) while LSD shows higher correlation with pesq (-0.85) and stoi (-0.74). VAD mismatch also shows moderate correlation with PESQ/STOI. Formant divergence and wav2vec confidence score show very low correlation with pesq and stoi.
In this work, we primarily focuses on phoneme-level artifacts, as these can introduce errors in downstream tasks such as ASR, so we focus on speech embeddings (wav2vec). 
\begin{figure}
    \centering
    \includegraphics[width=0.7\linewidth]{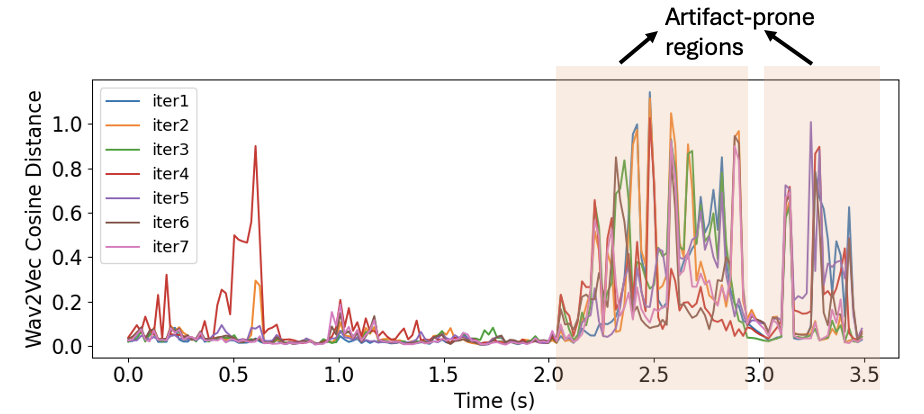}
    \caption{Cosine distance between clean and enhanced (multiple iterations) wav2vec embedding for the same example in Figure \ref{fig:multi_iter}}
    \label{fig:uncertainity}
\end{figure}
\subsection{Artifact Prediction using Uncertainty Analysis}

\begin{figure}[t]
\centering
\tiny
\setlength{\tabcolsep}{3pt}
\renewcommand{\arraystretch}{1.3}
\begin{tabular}{p{0.95\linewidth}}
\toprule
THEY DID NOT REPLACE IT WITH A CONVICTION FOR \textbf{CULPABLE HOMICIDE}  - clean file\\
THEY DID NOT REPLACE IT WITH A CONVICTION FOR \textcolor{red}{PHELPOVAL HOMICIDE} - iter 1 \\
THEY DID NOT REPLACE IT WITH HE CONVICTION FOR \textcolor{red}{FELFOBLE VOMECIDE} - iter 2 \\
THEY DID NOT REPLACE IT WITH THE CONVICTION \textcolor{red}{PORCOPOVAL HOMICIDE} - iter 3 \\
THEY DID NOT REPLACE IT WITH THE CONVICTION FOR \textbf{PULPABLE HOMICIDE} - iter 4 \\
THEY DID NOT REPLACE IT WITH A CONVICTION FOR \textcolor{red}{COPOVEL HOMECADE} - iter 5 \\
THEY DID NOT REPLACE IT WITH A CONVICTION FOR\textbf{ CULPABLE HOMECIDE} - iter 6\\
THEY DID NOT REPLACE IT WITH HE CONVICTION FOR \textcolor{red}{CULPRABAL VOMECIDE} - iter 7\\
\bottomrule
\end{tabular}
\caption{\footnotesize Multiple enhanced versions of the same noisy input. While some runs produce the correct phrase as shown in ASR transcriptions, others hallucinate spurious words (highlighted in \textcolor{red}{red}). This illustrates phoneme-level artifacts arising stochastically across diffusion samples. The wav2vec embedding variance for this example is shown in Figure \ref{fig:uncertainity}}
\label{fig:multi_iter}
\vspace{-0.2cm}
\end{figure}
As described earlier, typical SE metrics such as PESQ and STOI provide only limited insight into generative artifacts. To address this limitation, we use uncertainty analysis of speech embeddings. In diffusion-based SE, repeated runs with the same input can yield different outputs, since each sampling trajectory follows a stochastic path~\cite{lemercier2025diffusion}.
We observe that a fixed ASR model produces varying transcriptions for different enhanced samples (Figure \ref{fig:multi_iter}), likely due to artifacts generated under low-SNR conditions when the model is uncertain about its output.
In other words, hallucinations tend to occur in regions with high prediction variance, suggesting that the model’s score function experiences rapid gradient changes in those segments.
As shown in Figure \ref{fig:uncertainity}, the variance between the speech embeddings is high in artifact regions.
Using this observation, we can predict the probability of artifacts and their locations. 
As shown in algorithm \ref{alg:artifact_pred}, artifact regions are detected by computing the variance of wav2vec frame-level embeddings over time (Figure \ref{fig:uncertainity}). Later, we find average variance over time frames to get a single artifact score and compare it with threshold $\tau$.
\subsection{Artifact Reduction using Semantic Consistency}
As shown in Figure \ref{fig:multi_iter}, sometimes the model tends to hallucinate more, 
In some iterations, the enhanced output is very close to the clean output, resulting in less WER while other times the WER is very high. 
\begin{algorithm}[t]
\footnotesize
\caption{\small Artifact Prediction via Embedding Variance}
\label{alg:artifact_pred}
\begin{algorithmic}[1]
\Require Noisy input $y$, ensemble size $S$, threshold $\tau$, generative SE model $M$, encoder $E$
\Ensure Artifact score $a$, flag ${I}_{artifact}$
\State Generate $S$ enhanced candidates:
      $\{\hat{x}_1, \hat{x}_2, \ldots, \hat{x}_S\} \gets M(y)$
\State Extract frame-level embeddings for each candidate:
      $\{e_1(t), e_2(t), \ldots, e_S(t)\}$ using $E$
\State Compute per-frame variance across ensemble members:
      $v(t) = \operatorname{Var}\big(\{ e_1(t), e_2(t), \ldots, e_S(t)\}\big)$
\State Compute artifact score as mean variance across frames:
      \[
      a = \frac{1}{T}\sum_{t=1}^{T} v(t)
      \]
\If{$a > \tau$}
    \State Flag as \textbf{artifact-prone}, ${I}_{artifact} = 1$
\Else
    \State ${I}_{artifact} = 0$
\EndIf
\State \Return artifact score $a$, flag ${I}_{artifact}$
\end{algorithmic}
\end{algorithm}
\normalsize
\begin{algorithm}[t]
\footnotesize
\caption{\small Artifact Reduction via S-Ensemble}
\label{alg:artifact_reduct}
\begin{algorithmic}[1]
\Require Enhanced candidates $\{\hat{x}_1, \ldots, \hat{x}_S\}$, embeddings $\{e_1, \ldots, e_S\}$, reference embeddings ($e_{\text{clean}}$ or $e_{\text{noisy}}$)
\Ensure Final optimal enhanced output $\hat{x}$
\State Compute semantic similarity matrix:
      $C[i,j] = \text{corr}(e_i, e_j)$
\State Select candidate using one of the heuristics:
      \begin{itemize}
        \item \textbf{Ensemble centrality:} choose $\hat{x}_i$ with max average $C[i,:]$
        \item \textbf{Clean correlation:} choose $\hat{x}_i$ with highest correlation to $e_{\text{clean}}$
        \item \textbf{Noisy correlation:} choose $\hat{x}_i$ with highest correlation to $e_{\text{noisy}}$
      \end{itemize}
\State \Return selected output $\hat{x}$
\end{algorithmic}
\end{algorithm}
\normalsize
To solve this challenge, we aim to find the optimal enhanced output. We propose an ensemble inference strategy guided by semantic consistency. Given a noisy input, we generate $S$ enhanced outputs by sampling the diffusion model with different random seeds. Each enhanced waveform is passed through a self-supervised speech model (wav2vec) to extract speech embeddings that capture phonetic structure. We then compute the pairwise correlation between embeddings and select the sample that is most semantically consistent according to one of the following heuristics (Algorithm \ref{alg:artifact_reduct}):\\
\noindent \textbf{Clean correlation}: select the sample whose embedding is closest to the clean reference (used only for analysis). This is just for understanding the effectiveness of the approach, as clean audio is not available during inference.\\
\noindent  \textbf{Noisy correlation}: select the sample most correlated with the noisy input’s embedding, under the assumption that artifacts are not present in noisy input, they are generated by the diffusion modeling.\\
\noindent \textbf{Ensemble centrality}: select the sample whose embedding is most consistent with the rest of the ensemble, i.e., closest to the centroid of all samples.\\
These heuristics can be used in inference and do not influence training and model architecture.
The correlation with clean speech serves as an upper bound (the best we can achieve with this technique), while the noisy correlation and ensemble centrality provide practical, reference-free selectors.  
Semantic ensemble inference exploits the fact that hallucinations vary across diffusion runs, whereas genuine content remains consistent. By selecting the sample whose embedding is most semantically aligned, we favor phoneme-preserving outputs and suppress unstable artifacts. This requires no retraining, only multiple generations, with inference cost scaling by $S$. As shown later, reducing diffusion steps $N$ offsets this cost, making small ensembles ($S{=}3$–5) a practical solution for artifact-aware diffusion SE.
\begin{figure}
    \centering   \includegraphics[width=0.8\linewidth]{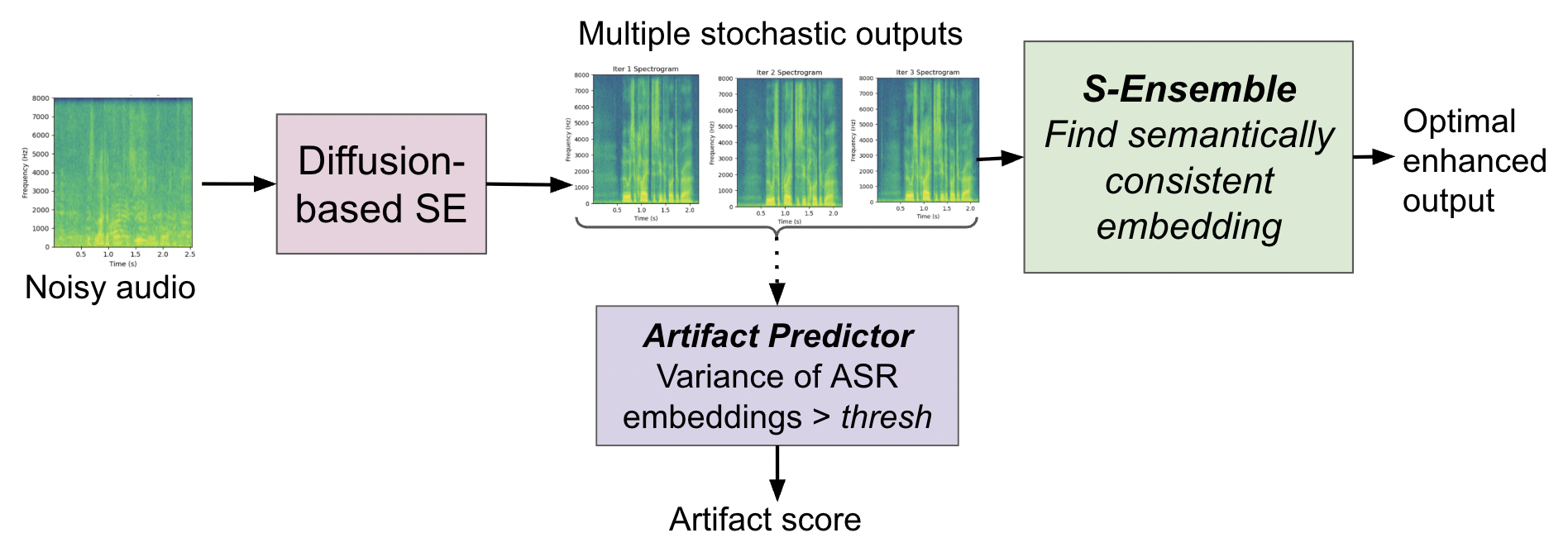}
    \caption{Proposed system framework}
    \label{fig:sys}
\end{figure}
\subsection{Effect of Reverse Diffusion Steps (N)}
\begin{figure}
    \centering    \includegraphics[width=1.0\linewidth]{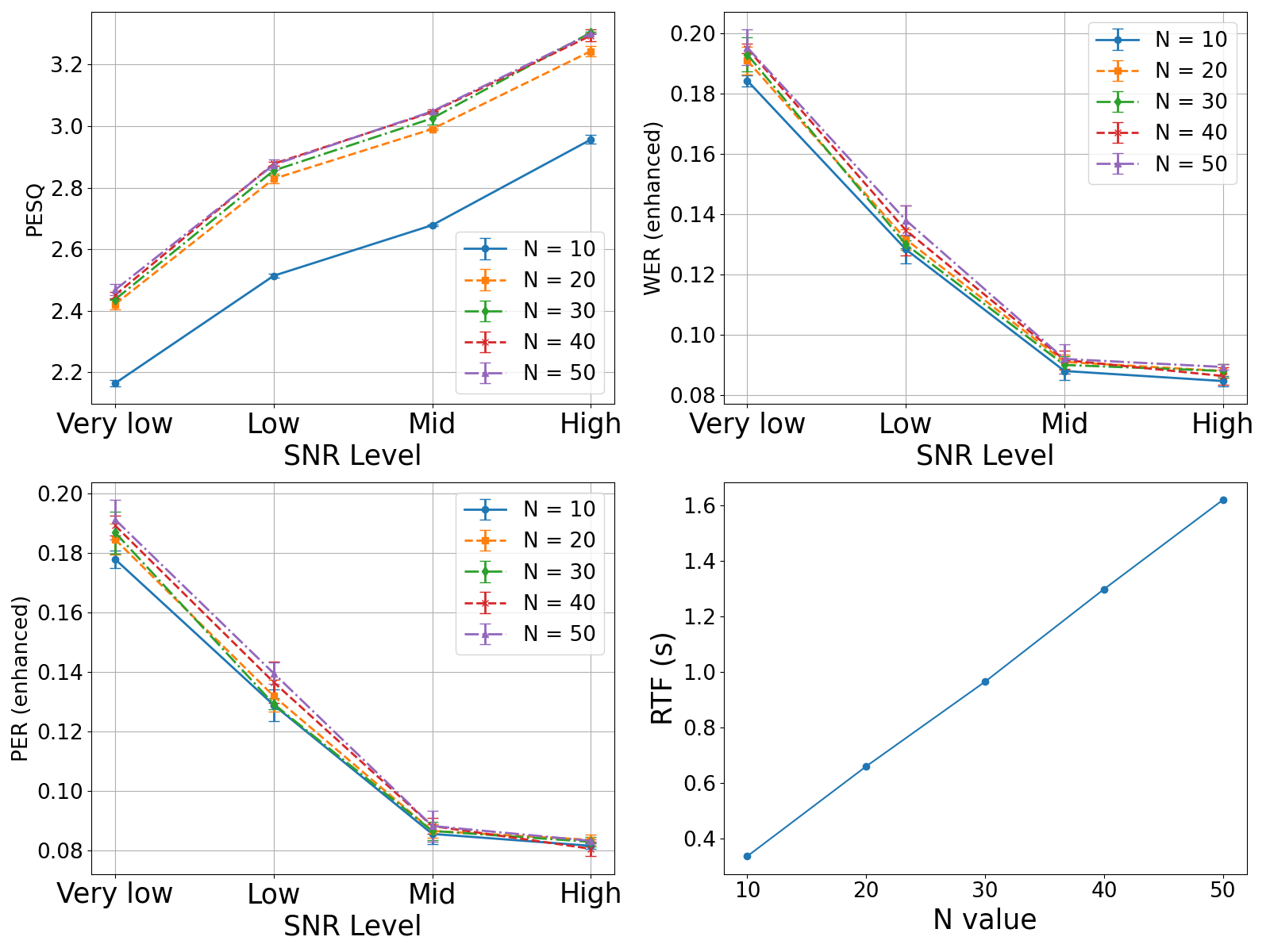}
    \caption{Effect of reverse diffusion steps on PESQ, WER, PER, and latency on VoiceBank-DEMAND test set}
    \label{fig:effectN}
\vspace{-0.2cm}
\end{figure}

We evaluate the impact of varying the number of reverse diffusion steps ($N$) during inference.
The plots in Fig.~\ref{fig:effectN} show PESQ, PER, WER, and real-time factor (RTF) across
different $N$ values (RTF is the inference latency defined in Section \ref{sec:setup}). As expected, larger $N$ values generally improve
perceptual scores (PESQ, STOI) but at the cost of significantly higher latency. In contrast,
WER and PER are more sensitive to $N$. First, at low SNR, there is high variance in WER/PER due to phoneme artifacts. Second, \textit{low $N$ values yields slightly lower WER/PER as less denoising steps means less chance of hallucinations}. But there could still be slight residual noise leading to low PESQ/STOI. 
Thus, lower N value often leads to less artifacts generation and linearly reduces the latency of diffusion models.
We try different N schedules for different SNR ranges to show that adaptive N can provide good performance-latency tradeoffs.
\vspace{-0.45cm}
\section{Evaluation}
\label{sec:evaluation}
\subsection{Experimental Setup}
\label{sec:setup}
We evaluate on publicly available VoiceBank+DEMAND (VBD). The VBD test set has 4 SNR levels: very low (-3 to 3 dB), low (4 to 8 dB), medium (9 to 13 dB), and high (13 to 18 dB). 
The low-SNR subsets is where artifacts are mostly found. From the low SNR set, we create a artifact test set which has the most prominent phoneme artifacts (noisy WER = 0.513 and clean WER = 0.115) through subjective listening. 
We use the pretrained SGMSE model trained on VBD train set to get enhanced output. 
For inference latency measurement, we use RTF which is the inference time for 1 sec audio file. 
RTF is measured on Lenovo ThinkStation P8 [GeForce RTX 4080]. 
For extracting embedding and measuring WER and PER, we use wav2vec 2.0 base model \cite{baevski2020wav2vec}. VBD test set has ground truth transcriptions for comparison. 

\subsection{Inference-Time Artifact Prediction}
\label{sec:artifact_prediction}

By computing wav2vec embeddings for $S$ independent samples and measuring the variance across them, we obtain an \emph{artifact score}. 
We extract $S$ enhanced outputs of same input using SGMSE model. 
We evaluate prediction accuracy and the variance thresholds required for different ensemble sizes $S$. 
In all cases, artifact detection achieved \textbf{100\%} accuracy, the optimal threshold varied slightly with $S$, ranging 
from $0.0067$ at $S{=}3$ to $0.0095$ at $S{=}7$. While larger ensembles produce more stable thresholds, even 
a small ensemble ($S{=}3$) is sufficient for perfect classification. This shows that variance across embeddings 
is a robust indicator of phoneme artifacts and can reliably flag uncertain regions during inference.
Thus, embedding variance across multiple diffusion iterations provides a lightweight and highly accurate inference-time detector for phoneme artifacts, enabling online artifact monitoring.


\subsection{Artifact Reduction with S-Ensemble}
\label{sec:semsemble}

We next evaluate our proposed ensemble inference strategy, which selects the most
semantically consistent sample among $S$ enhanced outputs. 
By extracting wav2vec embedding and applying correlation-based heuristics, we select the candidate
most likely to represent the true phonetic content.
Table~\ref{tab:semsemble_results} shows WER on the low-SNR VBD artifact test set. The baseline single-sample average WER across ten diffusion iterations was 0.438,
significantly worse than the WER of clean reference audio (0.115). \emph{S-Ensemble} substantially improves WER across
ensemble sizes $S{=}3$–10. Among the selection heuristics, correlation with the clean reference provides the best possible performance, achieving up to 36.3\% relative WER
reduction with $S{=}10$. The practical, reference-free heuristics also yield consistent gains:
most correlated improves WER by\textbf{ 14–16\% } across $S$, while most correlated with noisy
achieves 7–15\% relative improvement.
\begin{table}[t]
\centering
\footnotesize
\caption{ WER on low-SNR artifact test set using S-Ensemble with different ensemble sizes $S$.
Average single-sample WER of 10 iterations is 0.438. WER of clean set is 0.115 and noisy set is 0.513.}
\label{tab:semsemble_results}
\begin{tabular}{lccccc}
\toprule
\textbf{WER} & \textbf{S=3} & \textbf{S=5} & \textbf{S=7} & \textbf{S=9} & \textbf{S=10} \\
\midrule
Ensemble centrality & 0.373 & \textbf{0.368} & 0.373 & 0.373 & 0.373 \\
Clean correlation & 0.341 & 0.344 & 0.321 & 0.318 & \textbf{0.279} \\
Noisy correlation & 0.377 & 0.405 & 0.372 & 0.380 & 0.385 \\
\bottomrule
\end{tabular}
\end{table}
\normalsize
Thus, \textit{S-Ensemble} effectively mitigates phoneme artifacts by exploiting
semantic consistency across multiple diffusion outputs. 
This confirms that stochastic artifacts can be suppressed without retraining the model,
although at the cost of increased inference latency proportional to $S$.
\subsection{Adaptive Diffusion Steps ($N$)}
To address the latency overhead of ensemble inference, we evaluate the effect of varying the number of diffusion steps $N$. In the VBD testset, we have noise additions in 4 different SNR levels. For each of these SNR test set files, we use different N value to find the suitable N value for a given SNR. Table~\ref{tab:adaptiveN} compares fixed and adaptive schedules. 
Reducing $N$ lowers RTF substantially and slightly imporves WER. 
For example, $N{=}10$ reduces RTF by 65\%, improves WER by 4\%, with a modest 0.34 PESQ drop (-11.7\%). 
Since PESQ/STOI scores show similar values for different N ranging from 20 to 50, but drops at N=10 (See Figure \ref{fig:effectN}), adaptive schedules that allocate different $N$ values to different SNR can optimize the quality-latency tradeoff. By using lower $N=10$ value for high SNR and moderate $N=20$ for low SNR (to preserve PESQ score), we achieve a \textbf{27\%} RTF improvement with minimal quality loss.
\begin{table}[t]
\centering
\footnotesize
\caption{ Trade-off between RTF, PESQ, and WER for different $N$ schedules for different SNR sets [very low, low, mid, high]. Improvements are reported as relative percentage changes compared to fixed $N{=}30$.}
\label{tab:adaptiveN}
\begin{tabular}{lcccc}
\toprule
Heuristic & RTF & RTF $\Delta$ & PESQ $\Delta$ & WER $\Delta$ \\
\midrule
Fixed $N{=}30$ & 0.965 & -- & -- & -- \\
Fixed $N{=}10$ & 0.335 & \textbf{-65.3\%} & -11.7\% & -4.0\% \\
$[40,30,20,10]$ & 0.885 & -8.3\% & -3.1\% & -3.0\% \\
$[10,20,30,40]$ & 0.878 & -8.9\% & -3.4\% & -2.0\% \\
$[20,30,20,10]$ & 0.702 & -27.2\% & -3.4\% & -2.0\% \\
\bottomrule
\end{tabular}
\end{table}
\normalsize
\noindent Interestingly, we find that WER/PER vary more with $N$ than PESQ/STOI, confirming that artifacts are primarily semantic rather than acoustic. Lowering $N$ often reduces hallucinations by limiting stochastic updates.
This supports using SNR-aware adaptive $N$ for real-time applications.

\subsection{Optimizing $S$ and $N$}
Finally, we investigate whether a small number of diffusion steps combined with a small ensemble size can provide artifact reduction at no additional latency cost. Table~\ref{tab:n10s3} reports results of different heuristics for $N{=}10$ and $S{=}3$. While the baseline single-sample average WER is 0.428, the proposed technique reduces WER by 17\% depending on the selection heuristic
Notably, the most correlated with clean achieves a WER of 0.357, corresponding to a 16.5\% relative improvement. Since $N{=}10$ lowers inference time by 65\% compared to the default $N{=}30$, using $S{=}3$ iterations restores the runtime to that of the default setting while simultaneously reducing phoneme artifacts. This demonstrates that carefully balancing $N$ and $S$ can achieve both low-latency and artifact-aware enhancement. 

\begin{table}[t]
\centering
\footnotesize
\caption{Performance with $N{=}10$ and $S{=}3$ on low-SNR artifact test set. 
WER of clean set is 0.115 and noisy set is 0.513. PESQ for $N{=}30$ is 2.9 and $N=10$ is 2.57}
\label{tab:n10s3}
\begin{tabular}{lcc}
\toprule
\textbf{Method} & \textbf{WER} & \textbf{PESQ} \\
\midrule
Baseline avg (3 runs) N=10 & 0.428 & 2.577 \\
Ensemble Centrality (Most consistent) & 0.384 & 2.578 \\
Clean correlation & \textbf{0.357} & 2.575 \\
Noisy correlation & 0.371 & \textbf{2.578} \\
\bottomrule
\end{tabular}
\end{table}

%
\vspace{-0.4cm}
\section{Conclusion and Future Work}
\label{sec:conclusion}
We presented an artifact-centered study of diffusion-based SE. 
We predict artifacts using variance of speech embeddings and reduce artifacts via a semantic consistency-based ensemble inference strategy. We analyze the latency–artifact trade-off by varying reverse diffusion steps.
More broadly, our results suggest that artifact reduction can be framed as augmenting the diffusion posterior with a semantic prior. Wav2vec provides such a prior by embedding speech into a lower-dimensional space that captures phonetic meaning. By sampling multiple candidates from the generative model and ranking them according to semantic consistency in this embedding space, we effectively steer generation toward outputs that preserve meaning. Given a suitable encoder, semantic priors an be used in other generative enhancement tasks (dereverberation, bandwidth extension) or even beyond speech, e.g., in audio-to-text or image restoration.

\bibliographystyle{IEEEbib}
\bibliography{strings,refs}

\end{document}